\documentclass[journal,10pt,twocolumn]{IEEEtran}
\IEEEoverridecommandlockouts
\usepackage{amsmath,epsfig}
\usepackage{bm}
\usepackage{latexsym}
\usepackage{amsmath}
\usepackage{amssymb}
\usepackage{graphicx}
\usepackage[algonl,boxed]{algorithm2e}

\newtheorem{proposition}{Proposition}

\begin{document}

\title{Repeated Auctions with Learning for Spectrum Access in Cognitive Radio Networks}

\author{\authorblockN{Zhu Han$^\dag$, Rong Zheng$^\ddag$, and H. Vincent
Poor$^*$\\
\authorblockA{$^\dag$ECE Department, University of Houston, Houston, TX 77004\\
$^\ddag$CS Department, University of Houston, Houston, TX 77204\\
$^*$Princeton
University, Princeton, NJ 08544}}}

\maketitle

\begin{abstract}
In this paper, spectrum access in cognitive radio networks is modeled as a
repeated auction game subject to monitoring and entry costs.  For secondary
users, sensing costs are incurred as the result of primary users' activity.
Furthermore, each secondary user pays the cost of transmissions upon successful
bidding for a channel. Knowledge regarding other secondary users' activity is
limited due to the distributed nature of the network. The resulting formulation
is thus a dynamic game with incomplete information. In this paper, an efficient
bidding learning algorithm is proposed based on the outcome of past
transactions. As demonstrated through extensive simulations, the proposed
distributed scheme outperforms a myopic one-stage algorithm, and can achieve a
good balance between efficiency and fairness.
\end{abstract}

\section{Introduction}
Recent studies have shown that despite claims of spectral scarcity, the actual
licensed spectrum remains unoccupied for long periods of time~\cite{FCC}. Thus,
cognitive radio (CR) systems have been proposed~\cite{CR01} in order to efficiently
exploit these spectral holes. CRs or secondary users (SUs) are wireless devices
that can intelligently monitor and adapt to their environment, hence, they are
able to share the spectrum with the licensed primary users (PUs), operating
whenever the PUs are idle. Three key design challenges are active topics of
research in cognitive radio networks, namely, distributed implementation,
spectral efficiency, and the tradeoff between sensing and spectrum access.

Previous studies have tackled various aspects of spectrum sensing and
spectrum access. In \cite{CS00}, the performance of spectrum sensing, in terms of
throughput, is investigated when the SUs share their instantaneous knowledge of
the channel. The work in \cite{DT00} studies the performance of different
detectors for spectrum sensing, while in \cite{CS01} spatial diversity methods
are proposed for improving the probability of detecting the PU by the SUs.
Other aspects of spectrum sensing are discussed in \cite{CS02} and \cite{CS04}.
Furthermore, spectrum access has also received increased attention, e.g.,
\cite{SA00,SA01,SA02,SA03,SA04}.  In \cite{SA00}, a dynamic programming
approach is proposed to allow the SUs to maximize their channel access time
while taking into account a penalty factor from any collision with the PU. The
work in \cite{SA00} (and the references therein) establish that, in practice,
the sensing time of CR networks is large and affects the access performance of
the SUs. In \cite{SA01}, the authors model the spectrum access problem as a
non-cooperative game, and propose learning algorithms to find the correlated
equilibria of the game.  Non-cooperative solutions for dynamic spectrum access
are also proposed in \cite{SA02} while taking into account changes in the SUs'
environment such as the arrival of new PUs, among others. 

When multiple SUs compete for spectral opportunities, the issues of fairness
and efficiency arise. On one hand, it is desirable for an SU to access a
channel with high availability. On the other hand, the effective achievable
rate of an SU decreases when contending with many SUs over the most available
channel. Consequently, efficiency of spectrum utilization in the system
reduces. Therefore, an SU should explore transmission opportunities in other
channels if available and refrain from transmission in the same channel all the
time. Intuitively, diversifying spectrum access in both frequency (exploring
more channels) and time (refraining from continuous transmission attempts)
would be beneficial to achieving fairness among multiple SUs, in that SUs
experiencing poorer channel conditions are not starved in the long run.

The objective of this paper is to design a mechanism that enables fair and
efficient sharing of spectral resources among SUs.  We model spectrum access in
cognitive radio networks as a repeated auction game with entry and
monitoring costs.  Auctioning the spectral opportunities is carried out
repeatedly. At the beginning of each period, each SU that wishes to participate in
the spectrum access submits a bid to a coordinator based on its view of the
channel and past auction history. Knowledge regarding other secondary users'
activities is limited due to the distributed nature of the network. The
resulting formulation is thus a dynamic game with incomplete information. The
bidder with the highest bid gains spectrum access. Entry fees are charged for all bidders who participate in the auction
irrespective of the outcome of the auction.  An SU can also choose to stay out
(SO) of the current round, in which case no entry fee is incurred. At the end
of each auction period, information regarding bidding and allocation are made
available to all SUs, and in turn a monitoring fee is incurred.

To achieve efficient bidding, a learning algorithm is proposed based on the
outcome of past transactions. Each SU decides on local actions with the
objective of increasing its long-term cost effectiveness. As demonstrated
through extensive simulations, the proposed distributed scheme outperforms a
myopic one-stage algorithm where an SU always participates in the spectrum
access game in both single channel and multi-channel networks.

A comment is in order on the feasibility of such an auction-based
approach to spectrum access in practice. Due to commercial and industrial
exploitation and different stake holders' interests, the functional
architectures and cognitive signaling schemes are currently under discussion
within standardization forums, including IEEE SCC 41 and ETSI TC RRS
(Reconfigurable Radio Systems). Cognitive pilot channel (CPC) has gained attention
as a potential enabler of data-aided mitigation techniques between secondary
and primary communication systems as well as a  mechanism to support
optimized radio resource and data management across heterogeneous networks. In
CPC, a common control channel is used to provide the information corresponding
to the operators, Radio Access Technology and frequencies allocated in a given
area. We can thus leverage the intelligence of the CPC coordinator and the
control channel to solicit bidding and broadcast the outcome of auctions.

The main contributions of this paper are:
\begin{enumerate}
\item We have formulated the spectrum access problem in cognitive radio networks
as a repeated auction game.
\item A distributed learning algorithm is proposed for single-channel networks,
and a non-regret learning algorithm is investigated for multi-channel networks.
\end{enumerate}
The rest of the paper is organized as follows. In Section~\ref{sec:model}, the
system model and terminology are introduced. Mechanism design of the repeated
auction with learning is presented in Section~\ref{sec:algorithm}.  Simulation
results are given in Section~\ref{sec:simulation} followed by conclusions and a
discussion of future work in Section~\ref{sec:conclusion}.
\section{Physical layer and System Model}
\label{sec:model}
We consider a cognitive radio network consisting of $K$ channels to be occupied
by $N$ SUs who compete repeatedly for access to the
channels at each discrete time $t$. At time $t$, the $i^{th}$ SU can reasonably
estimate its channel rate $\theta_{i,k}^t$ in the $k^{th}$ channel while having no
knowledge of that of other SUs.  In other words, each SU has imperfect
information.  We assume that both $N$ and $K$ are known to all $SUs$.  The
primary user's activity follows Bernoulli
distribution, i.e., at time $t$, the $k$th channel is  occupied with
probability $\Theta_k$ at time $t$. Without loss of generality, all secondary
users use a common transmit power $P_0$ with a thermal noise level $\sigma^2$ at the basestation.
The channel gain for the $i^{th}$ secondary user is assume to be
$G_ih_{i,k}^t$, where $G_i$ is the propagation path loss and $h_{i,k}^t$
follows the Rayleigh fading distribution. The rate for the $i^{th}$ user on the
$k^{th}$ channel at time $t$ can be written as
\begin{equation}
\theta_{i,k}^t=W\log_2\left(1+\frac{G_ih_{i,k}^t}{\sigma^2}\right),
\end{equation}
where $W$ is the bandwidth for each channel. All channels are assumed to
have the same bandwidth for ease of exposition.

\begin{figure}[htb]
 \centerline{\epsfig{figure=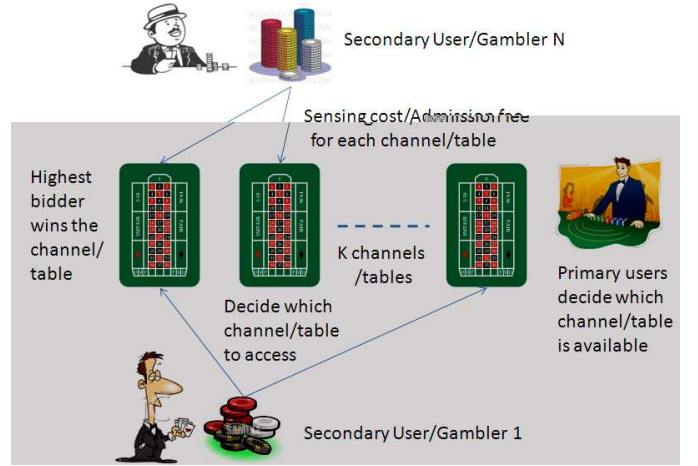,width=3.5in}}
\caption{Illustration of the System Model}
 \label{fig:system_model}
\end{figure}

At time $t$, an SU may incur two types of costs, namely, the cost of accessing
$c_t$, which accounts for the energy expenditure needed for spectrum access; and
the cost of monitoring $e_t$, which is the cost of sensing and subscribing to
the control channel (e.g., CPC) to obtain information regarding past auctions.
The spectrum access among SUs is modeled as a repeated auction. The access
cost, also called the entry free is charged only when the user decides to
participate in the auction at time $t$.  On the other hand, SUs always need to
pay for the monitoring cost regardless of their decisions. At the
beginning of a slot $t$, an SU chooses whether to stay out (SO) or participate
in spectrum access. If the latter option is chosen, the SU (or bidder) sends a
confidential message to the coordinator (or auctioneer) containing its bid. Let the bid submitted
by SU $i$ be $\textbf m_{i}^t$, which is a $K\times 1$ vector with component
$m_{i,k}^t$ for the $k^{th}$ channel.  We define the set of actions of user
$i$'s opponents as
\begin{equation}
\textbf m_{-i}^t=\{\textbf m_{j}^t|j\in N\backslash i\}.
\end{equation}

The cost incurred is thus $$e_t+c_t 1(m_{i,k}^t\neq SO),$$ where $1(\cdot)$ is the
indicator function.

In each round, the bidder with the maximum bid wins and is granted  the
spectrum access opportunity. A payment is incurred accordingly. There are two
key differences compared to existing spectrum access models, where upon
sensing an idle channel, all SUs contend for spectrum access. First, an SU may
choose to stay out if participating in the spectrum access game is deemed
unfavorable (because of low data rate or large number of contending SUs).
Second, the transmission opportunity in each available channel is granted only
to a single SU after the auction. Therefore, no further contention will occur.
Each SU is assumed to follow a symmetric strategy based on its local state and
information learned.

Figure \ref{fig:system_model} gives an illustration of the system model.
An analogy can be drawn in casinos, in
which different gamblers try to select which table to play and how much to bet.  Each
secondary user shall decide which channel to sense and bid for, and how
much the bid should be. These two issues will be addressed in the following
sections for single-channel (i.e., $K=1$) and multi-channel (i.e., $K>1$)
cognitive radio networks, respectively.


\section{Repeated Auction with Learning}
\label{sec:algorithm}
In this section, we investigate the spectrum access problem among multiple SUs
in cognitive radio networks.  We first discuss the auction mechanism and then
define the utility function.  Finally, an efficient bidding mechanism in
repeated auctions with learning is proposed.

\subsection{Mechanism}
Recall from Section~\ref{sec:model} that at the beginning of a slot $t$, an SU
decides to either place a bid, or stay out and monitor the results. Based on the
SUs' actions, the allocation strategy at time $t$ for channel $k$ can be written as
$$X_t(k) \stackrel{\Delta}{=} \{\chi_{1, k}^t, \ldots,\chi _{n,k}^t| \chi_{i,k}
\in \{0, 1\}\wedge \sum_{i}{\chi_{i,k}^t} = 1\}, \forall k.$$ If SU $i$ chooses to stay out, then
its allocation equals zero, i.e.,
\begin{equation}
\chi_{i,k}^t(m_{i,k}^t=SO)=0.
\end{equation}
The SU with the highest bid would win the right to access the channel, i.e.,
\begin{equation}
\chi_{i,k}^t(m_{i,k}^t\neq SO)=\left\{
\begin{array}{ll}
1, & m_{i,k}^t>m_{j,k}^t, \forall j\neq i;\\
& \mbox{and primary user does not exist}\\
0, & \mbox{Otherwise}.
\end{array}
\right.
\end{equation}
The winner would pay
\begin{equation}
p_{i,k}^t=\left\{
\begin{array}{ll}
0, & m_{i,k}^t=SO \mbox{ or } \exists j, m_{j,k}^t>m_{i,k}^t;\\
& \mbox{and primary user exists}\\
\psi_{i,k}^t, & \mbox{Otherwise}
\end{array}
\right.
\end{equation}
for its bid, where
\begin{equation}
\psi_{i,k}^t=\left\{
\begin{array}{ll}
m_{i,k}^t, & \mbox{First Price Auction};\\
\max(\textbf m_{-i,k}^t), & \mbox{Second Price Auction}.
\end{array}
\right.
\end{equation}

For ease of presentation, a second price auction is assumed in the remaining
discussion of the paper. The auction mechanism can be written as follows:
\begin{enumerate}
\item SU $i$ observes its current valuation (rate) $\theta_{i,k}^t$;

\item SU $i$ decides $m_{i,k}^t$;

\item The mechanism implements $\chi_{i,k}^t$ and $\psi_{i,k}^t$; and

\item SU $i$ observes  $\chi_{i,k}^{t}$ and $p_{i,k}^t$.

\end{enumerate}

Mechanisms and results for ``one-shot" auction with and without entry fees have
been well established in the literature~\cite{RePEc:nwu:cmsems:1096}. Typically, a
symmetric and known strategy is assumed. In our formulation, at each stage of
the auction, an SU decides on its action according to the bidding history
monitored thus far. The number of participants varies from stage to stage
depending on the SU's valuation and its knowledge regarding other players.
\subsection{Utility Function}
To assess the expenditure in the course of the game, we define the accumulated cost
for SU $i$ at time $t$ as
\begin{equation}
c_{i}^t(h^{t-1}_i)=\sum_{k=1}^K \sum_{\tau=1}^{t-1} \left[ p_{i,k}^\tau\chi_{i,k}^\tau+c_t 1(m_{i,k}^\tau \neq 0)+ e_t\right],
\end{equation}
where $h^{t-1}_i$ is the bidding history observed by SU $i$ up to time $t$. The
cost includes payment for the spectrum access opportunity, entry and monitoring
fees over the history and across the different channels.

The accumulated reward of SU $i$ is given by
\begin{equation}\label{eqn:reward}
r_{i}^t(h^{t-1}_i)=\sum_{k=1}^K  \sum_{\tau=1}^{t-1} \theta_{i,k}^\tau \chi _{i,k}^\tau.
\end{equation}

The utility is thus defined as the accumulated reward over the
total cost, i.e.,

\begin{equation}\label{eqn:utility}
\gamma_{i}^t(h^{t-1}_i)=\frac{r_{i}^t(h^{t-1}_i)}{c_{i}^t(h^{t-1}_i)}.
\end{equation}

The utility function is essentially the revenue to cost ratio of the SU's actions
over time. An SU will try to maximize its utility. Intuitively, when an SU's
valuation is low compared with others, it is beneficial for the SU to stay out
so that the entry cost is not incurred unnecessarily. On the other hand, staying
out all the time leads to zero accumulation of revenue and starvation of the
SU, and thus should be avoided.  Optimizing (\ref{eqn:utility}) is difficult
even in a centralized manner due to the large decision space. Therefore,
distributed heuristic learning algorithms are warranted   to determine
$m_{i,k}^t$ at each SU individually.

At time $t$, an SU decides whether to participate in the auction and if so, its
bid.  If SU $i$'s decision is to participate (or $m_{i,k}^t \neq SO$), it can be
proved straightforwardly that SU $i$'s dominating strategy is to bid its own
valuation in the second price auction. More formally, we have
\begin{proposition}
The equilibrium of the repeated auction with utility function \eqref{eqn:utility}
consists of each bidder using the following strategy at time $t$:
\begin{equation}
m_{i,k}^t = \left\{\begin{array}{cc} SO & f(\theta_{i,k}^t, h_i^i{t-1}) \ge 0 \\
				    \theta_{i,k}^t & else, \\
		 \end{array}\right.
\end{equation}
\end{proposition}
where $f(\theta_{i,k}^t, h_i^{t-1})$ is a function of SU $i$'s current valuation and bidding
history. The above strategy implies a thresholding criterion for participating in the
game. The form of $f(\theta_{i,k}^t, h_i^{t-1})$ differs in the single channel and
multi-channel scenarios, and will be discussed in more detail in subsequent
sections.
\subsection{Repeated Auction in a Single Channel}
When there is only a single channel, we can drop $k$ in the notation. An SU stays
out of bidding if it deems that participation is likely to reduce its
payoff. Formally,
$m_{i}^t = SO$, if
\begin{equation}
\gamma_{i}^{t+1}(\theta_i^t,h^{t-1}_i:m_{i}^t = SO)  \ge
\mathbb{E}_{\theta_{-i}^t}\left(\gamma_{i}^{t+1}(\theta_i^t,h^{t-1}_i:m_{i}^t =
\theta_i^t)\right).
\label{eq:thresh}
\end{equation}

In \eqref{eq:thresh}, the expectation is taken over all possible valuations of
SU $i$'s opponents.

In the first auction, no past history is available. The same thresholding
function is applied at each SU under the assumption that the valuations of
SUs are independent and identically distributed  with cumulative distribution
function (CDF) $F(\cdot)$ and probability density function (PDF) $f(\cdot)$.
Therefore, the CDF and PDF of the second largest valuation among $N$ users are $G(y)
= F(y)^{N-1}$ and $g(y) = (N-1)F(y)^{N-2}f(y)$, respectively. Let $r_i^1 = c_i^1 = 1$, for
all $i$. The strategy for the first auction is stated as follows.
\begin{proposition}
In the first auction, $m_i^1 = SO$ if and only if $\theta_i^1 < \overline{c}$, where
$$\overline{\theta}G(\overline{\theta}) = \frac{e_1}{1+c_1}.
$$
\label{prop:first}
\end{proposition}
\begin{proof}
Since $\overline{c}$ is the lowest valuation of any SU to participate in the
auction, only when all other SUs have a valuation less than $\overline{\theta}$
will SU $i$ with valuation $\overline{\theta}$ win the auction. Therefore,
$$\gamma_{i}^{1}(\theta_i^1:m_{i}^1 = SO)  = \frac{1}{1+e_1},$$ and
$$\mathbb{E}_{\theta_{-i}^1}\left(\gamma_{i}^{1}(\theta_i^1:m_{i}^1 =
\overline{\theta})\right) = \frac{1+G(\overline{\theta})\overline{\theta}}{1+e_1+c_1}.$$
To satisfy \eqref{eq:thresh}, we have $$\overline{\theta}G(\overline{\theta}) = \frac{e_1}{1+c_1}.
$$
\end{proof}

Direct evaluation of \eqref{eq:thresh}  is difficult after the first
auction. This is because the accumulated reward, cost and current valuation are
only available to each SU individually (although the auctioneer provides the
information regarding the highest bid and associated payment at the end of each
stage). Next, we introduce a simple heuristic to approximate the right hand
side of \eqref{eq:thresh}. SU $i$ maintains a private threshold value
$\overline{\theta_i}$, initiated according to Proposition~\ref{prop:first}. At time $t$, SU
$i$ updates
$\gamma_{i}^{t}(\theta_i^t:m_{i}^t = SO) = \frac{r_{i}^t}{c_{i}^t + e_t}$.
Furthermore,
$$\mathbb{E}_{\theta_{-i}^t}\left(\gamma_{i}^{t}(\theta_{i}^t:m_{i}^t = \theta_i^t)\right) \approx \frac{r_{i}^t + \theta_i^t-\overline{\theta_i}}{c_{i}^t + e_t + c_t}.$$ SU $i$'s action is thus,
\begin{equation}
m_{i}^t = \left\{\begin{array}{cc} SO & \frac{r_{i}^t}{c_{i}^t + e_t} > \frac{r_{i}^t + \theta_i^t-\overline{\theta_i}}{c_{i}^t + e_t + c_t} \\
				    \theta_{i}^t & else \\
		 \end{array}\right..
\end{equation}
At the end of stage $t$, the SU obtains the largest bid and associated payment.
If SU $i$ chooses to stay out, but the payment of the winner is less than
$\overline{\theta_{i}}$, its $\overline{\theta_{i}}$ is set to the payment amount.
Otherwise, $\overline{\theta_{i}}$ remains the same. On the other hand, if SU
$i$ participates in the auction but either loses the auction or is required to
make a payment higher than  $\overline{\theta_{i}}$, its
$\overline{\theta_{i}}$ is set to the payment amount. To avoid fluctuation of
$\overline{\theta_{i}}$ estimates, a moving average of old and new values
can be applied.

The above mechanism is summarized in Algorithm~\ref{algo_singlechannel}.

\begin{algorithm}
\label{algo_singlechannel}
\SetKwData{Left}{left}
\SetKwInOut{Input}{input}
\SetKwInOut{Output}{output}
\SetKwInOut{Init}{init}
\SetKwFor{For}{for}{do}{endfor}
\caption{Strategy in single-channel spectrum access}
\Input{Number of SU's $n$, monitoring $e_t$ and entry fee $c_t$ at time $t$}
\Init{Set $\overline{\theta_i} = \overline{\theta}$ s.t., $\overline{\theta}G(\overline{\theta}) = \frac{e_1}{1+c_1}$}
\BlankLine
{$a = \gamma_{i}^{t}(\theta_{i,t}:m_{i}^t = SO)$}\;
{$b = \frac{r_{i}^t + \theta_i^t-\overline{\theta_i}}{c_{i}^t + e_t + c_t}$}\;
\If{$a > b$ or PU detected} {
$m_{i,t} = SO$;
}
\Else{
$m_{i,t} = \theta_{i,t}$
}
Let the maximum payment at stage $t$ be $p_m(t)$\;
\If{$m_{i,t} = SO$ and $p_m(t) < \overline{\theta_i}$} {
$\overline{\theta_i} = p_m(t)*\alpha + \overline{\theta_i}*(1-\alpha)$
}
\If{$m_{i,t} \neq SO$ and ($\chi_{i,t} = 0$ or $p_m(t) \ge \overline{\theta_i}$)} {
$\overline{\theta_i} = p_m(t)*\alpha + \overline{\theta_i}*(1-\alpha)$
}
\end{algorithm}

\subsection{Non-Regret Algorithm for the Multi-channel Case}\label{sec:multi-channel}

%
%
%
%

In this section, we will address the spectrum access problem in multi-channel
networks. A class of algorithms called regret-matching
\cite{Hart_Mas-Colell00} is explored. The resulting stationary solution of the
learning algorithm exhibits no regret by setting the probability of a
particular action proportional to the ``regrets" for not having played other
actions. In particular, for any two
distinct actions $\textbf{m}_i^t\neq \bar {\bf  m}_i^{t}$ at every time $t$, the regret of SU $i$ at time $t$ for not
playing $\bar{\bf m}_i^t$ is
\begin{equation}\label{eqn:Regret2}
\mathbb{R}_i^t(\textbf{m}_i^t,\bar {\bf  m}_i^{t}):=\max\{D_i^t(\textbf{m}_i^t,\bar {\bf  m}_i^{t}),0\},
\end{equation}
where
\begin{equation}\label{eqn:Regret1}
D_i^t(\textbf{m}_i^t,\bar {\bf  m}_i^{t})=\frac{1}{\nu}\sum_{t-\nu \leq \tau < t}
(r^t_i(\bar {\bf  m}_i^{t},\bar {\bf  m}_{-i}^{t})-r^t_i(\textbf{m}^t_i,\textbf{m}^t_{-i})),
\end{equation}
with$\nu$  denoting the size of time window.  $D_i^t(\textbf{m}_i^t,\bar {\bf
m}_i^{t})$ has the interpretation of average payoff that SU $i$ would have
obtained, if it had bid $\bar {\bf  m}_i^{t}$ every time in the past
instead of choosing $\textbf{m}_i^t$. The expression
$\mathbb{R}_i^T(\textbf{m}_i^t,\bar {\bf  m}_i^{t})$ can be viewed as a measure
of the average regret. In the context of spectrum access in multi-channel
networks, the alternative actions correspond to participating in the auction
game in different channels\footnote{Each user decides which channel to bid on, and then uses the value $\theta_{i,k}^t$ for bidding that channel.}. The probability $\textbf P_i^t$ for SU $i$ to join
auctions in  channel $k$ is a linear function of the regret. Here $P_i^t$ is a
$K$-by-$1$ probability vector with $\sum_{k=1}^K P_{i,k}^t=1$. Define $\textbf
1 (\textbf m_i^t)$ as the indication vector for whether or not SU $i$ competes
in the $k$th channel. The detailed regret-matching algorithm is given in Algorithm
\ref{table:regretmatching}. The complexity of the algorithm is $O(K)$ and can
be implemented distributively. Furthermore, its convergence has been
proved in the literature \cite{Hart_Mas-Colell00}. Once SU $i$ chooses the
channel to access, its action is decided by Algorithm~\ref{algo_singlechannel}.
Note that even though an SU can access only one channel at a time, the bidding
history on all data channels is made available through the control channel to
all SUs.


\begin{algorithm}
\SetKwData{Left}{left}
\SetKwInOut{Input}{input}
\SetKwInOut{Output}{output}
\SetKwInOut{Init}{init}
\SetKwFor{For}{for}{do}{endfor}
\caption{Non-regret learning algorithm for  the multi-channel case}\label{table:regretmatching}
\Init{The probability of SU $i$, $\textbf P_{i}^1$ is set arbitrarily}
\ForEach{$t=1,2,3,...$} {
    Find $D_i^t(\textbf{m}_i^t,\bar {\bf  m}_i^{t})$ as in (\ref{eqn:Regret1})\;
    Find average regret $\mathbb{R}_i^t(\textbf{m}_i^t,\bar {\bf  m}_i^{t})$ as in
    (\ref{eqn:Regret2})\;
    $\textbf P_i^{t+1}(\bar {\bf m}_i^t ) =\frac{1}{\kappa}\mathbb{R}_i^t(\textbf{m}_i^t,\bar {\bf  m}_i^{t})\textbf 1 (\bar{\bf m}_i^t),
    \forall \textbf{m}_i^t\neq \bar {\bf  m}_i^{t}$\;
     $\textbf P_i^{t+1}(\textbf m_i^t)=\left[ 1-\sum_{\bar {\bf m}_i^t\neq
    \textbf{m}^t_i}\textbf P_i^{t+1}(\bar {\bf m}_i^t )\right] \textbf 1 (\textbf m_i^t) $,\
     where $\kappa$ is a certain constant that is
     sufficiently large;
  }
\end{algorithm}

%
%
%

\begin{figure}[htb]
 \centerline{\epsfig{figure=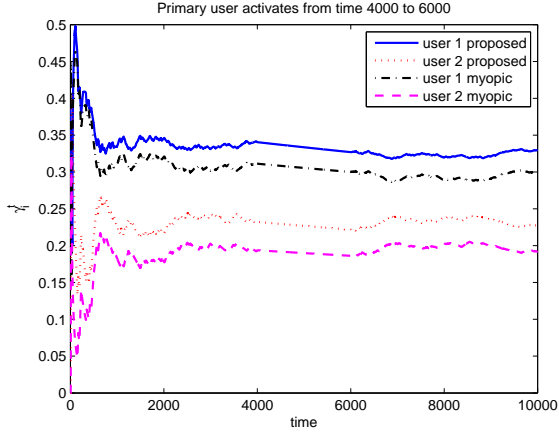,width=8cm}}
\caption{Convergence of repeated auction games under the proposed and myopic schemes. 2 SUs, 1 channel}
 \label{fig:r_vs_t_2user}
\end{figure}

\section{Simulation}\label{sec:simulation}

In this section, we investigate the performance of the proposed schemes by
simulations. We construct a network of dimensions $100$m-by-$100$m, in which the SUs are
randomly placed. All SUs transmit to a base station at a fixed location $1000$m away from the center of the network. The propagation loss exponent is set to be $3$. The common transmission power level
of all SUs is set to be 100mW and the noise level at -90dbm. A unit bandwidth
is assumed with frame length at 100$\mu$s, and Doppler frequency 100Hz. We set $\alpha=0.05$ and $\nu=10$. The
proposed schemes are compared to a myopic scheme, in which SUs always participate
in the auctions.

\begin{figure}[htb]
 \centerline{\epsfig{figure=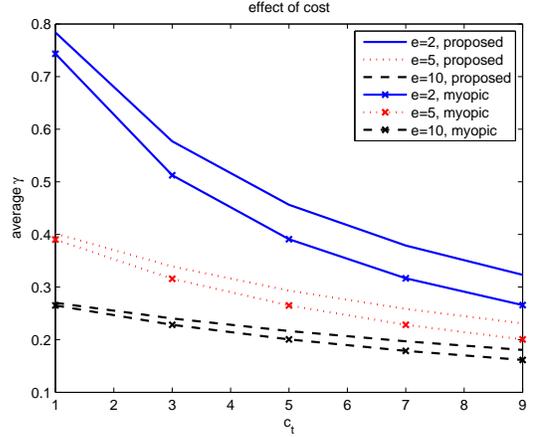,width=8cm}}
\caption{Effects of monitoring and entry costs. 2 SUs}
 \label{fig:r_vs_ct_e_2user}
\end{figure}

\paragraph*{Single channel} First, we consider a simple 2-user case to understand
the convergence of the proposed algorithm. Figure \ref{fig:r_vs_t_2user} shows
a snapshot of the change of the utility function $\gamma_{i,t}$ for user $i$ over
time. The entry and monitoring fees are fixed at $c_t=10$ and $e_t=1$,
respectively. From the figure, we can see that the proposed scheme converges
quickly and then tracks the changes in the channel.  Furthermore, compared to the
myopic scheme in which the SUs always bid, the utilities attained are higher for
both users. This is because the SUs can decide whether to bid or not based on its
valuation and outcome of past auctions. Between time $4000$ and $6000$, a primary
user is active, and all SUs stay out of the auction but still pay for the
monitoring fee. The average value of gamma decreases during that period of time.  After
the primary user stops transmitting, the auction game resumes.  Since the
effects of a primary user are very predictable,  in the remaining simulations we
assume the primary user is always idle.

\begin{figure}[htb]
 \centerline{\epsfig{figure=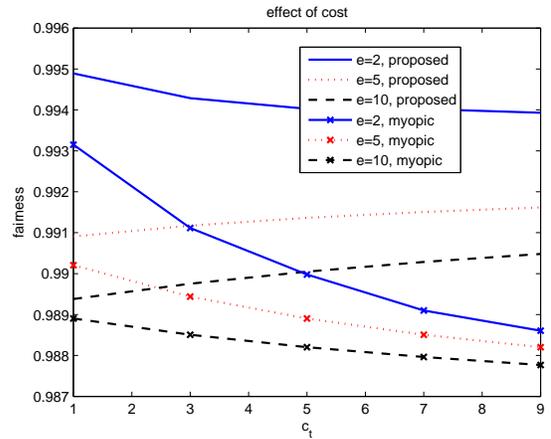,width=8cm}}
\caption{Fairness achieved by SUs}
 \label{fig:fairness}
\end{figure}

In Figure \ref{fig:r_vs_ct_e_2user}, we demonstrate the effects of entry and
monitoring costs on performance. The proposed scheme is shown to achieve better
performance in all cases and the average gain is up to 15\%. We can see that
when the monitoring fee is fixed, as the entry fee ($c_t$) increases, the average
utility decreases. This is expected as it is more expensive to participate in the
auction. Furthermore, the gap between the utilities attained by the proposed
scheme and the myopic scheme also increases. This is because in the
proposed scheme SUs are selective and would participate in the auction only if
they are likely to win. The myopic scheme would incur high losses in revenue as
the result of a higher entry cost.

In Figure \ref{fig:fairness}, we show the fairness achieved in the proposed and
myopic schemes. We adopt Jain's fairness index~\cite{jain84}:  $$F =
\frac{(\sum_{i=1}^N{\gamma_i})^2}{N\sum_{i=1}^N{\gamma_i^2}}.$$ Clearly, $F$ is between
$0$ and $1$.  The larger the value, the better the fairness is. We can see that the
proposed scheme results in fairer resource allocation compared with the myopic
bidding scheme. As the entry cost increases, the fairness of the myopic scheme also
decreases. This is because users experiencing worse channels are repeatedly
penalized by losing the game and paying entry fees. In comparison, when $e =
10$ and $e=5$, the proposed scheme achieves slightly better fairness as the
entry cost $c$ increases.

\begin{figure}[htb]
 \centerline{\epsfig{figure=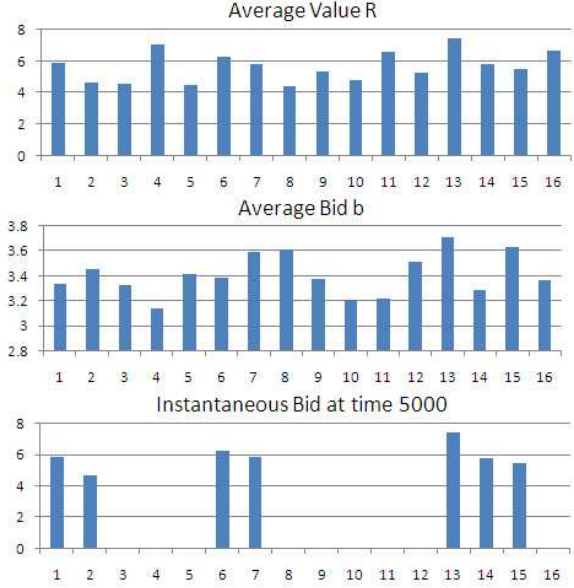,width=8cm}}
\caption{Valuation and bids of SUs}
 \label{fig:multiuser}
\end{figure}

\begin{figure}[htb]
 \centerline{\epsfig{figure=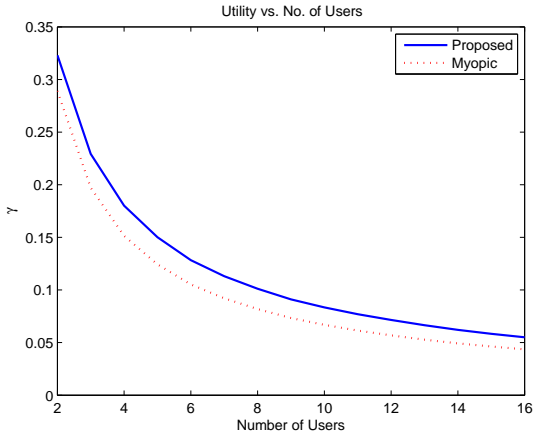,width=8cm}}
\caption{Utility achieved as the number of SUs changes}
 \label{fig:gamma_no_users}
\end{figure}

In the next set of experiments, we set the number of SUs to be $N=16$.  Figure
\ref{fig:multiuser} ($c_t=e_t=5$) compares the SUs' average valuation $R$, average
bid $b$ and instantaneous bid at time $5000$, respectively. Several
observations are in order.  First, the users with a higher
average value generally agree with users with a higher average bid though not
always. This is because the average bid also includes the case in which an SU stays
out (treated as a zero bid). Second, as expected, not all the users
are bidding in each slot; only the users with low cost and high chance of
winning would participate.

\begin{figure}[htb]
 \centerline{\epsfig{figure=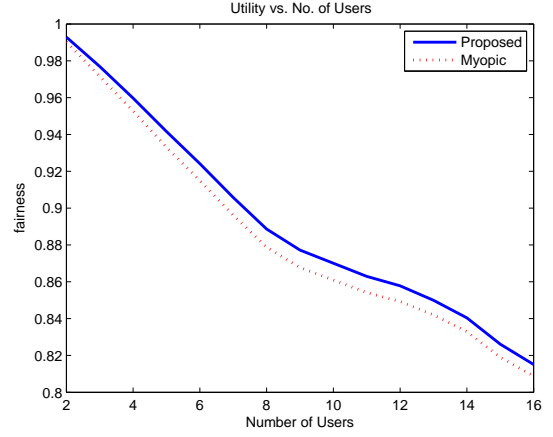,width=8cm}}
\caption{Fairness achieved as the number of SUs changes}
 \label{fig:fairness_no_users}
\end{figure}

\begin{figure}[htb]
 \centerline{\epsfig{figure=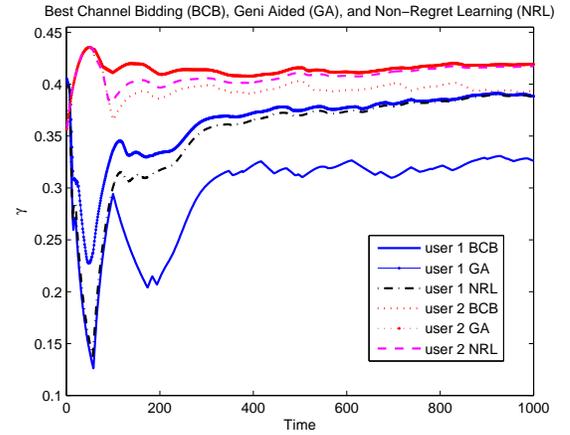,width=8cm}}
\caption{Utility in two-SU two-channel networks under different algorithms}
 \label{fig:multi_channel}
\end{figure}

In Figure \ref{fig:gamma_no_users} and Figure \ref{fig:fairness_no_users}, we
show the utility and fairness as a function of the number of SUs varying from 2
to 16. The costs are set as $c_t=10$ and $e_t=1$, respectively. We can see that
as the number of SUs increases, the utility decreases due to limited radio
resources. The fairness also decreases since there might be more chances for
users to dominate when the number of users is large. The proposed scheme has
better performance in both utility and fairness, compared with the myopic
scheme. The gain in utility ranges from 12\% to 25\%.


\paragraph*{Multiple channels}

In this set of experiments, we study the performance and convergence of a
two-user two-channel case. The parameters are set as $c_t=5$, $e=5$,
and $\nu=10$.  Three schemes are compared,
namely, Best Channel Bidding (BCB), Geni Aided (GA), and Non-Regret Learning
(NRL). In BCB, the users select to bid on the channel with the highest channel
gain. In GA, a Geni tells the SUs not to bid on the channel that they would not
win and instead to bid on the other channels. The GA solution is thus the performance
upper bound for practical systems. We can see that the BCB has the worst
performance since the SUs might bid on the same channel while the other
channels are vacant. The proposed NRL solution on the other hand, performs
closely to the GA solution, and can be easily implemented in a distributed
manner.

\section{Conclusions}\label{sec:conclusion}
In this paper, we have investigated the problem of spectrum access in single and
multi-channel cognitive radio networks. A repeated auction based framework has been
adopted. In single-channel spectrum access, SUs selectively participate in the
auction based on their valuation and past auction history. This scheme has been shown
to outperform a myopic scheme in which SUs always compete. In multi-channel
networks, a non-regret approach has been proposed. Its performance has been shown to be
significantly better than a naive greedy solution and come close to that of the Geni
aided solution. As future work, we plan to improve on the convergence speed and
optimality of the proposed scheme. Also of interest is the study of robust
mechanisms for situations in which the monitored information may be inaccurate.

\section*{Acknowledgments}
This research was supported in part by the Air Force Office of Scientific
Research under Grant FA 9550-08-1-0480 and the National Science Foundation under
Grant CNS-0832084.

\bibliographystyle{abbrv}

\end{document}